\documentclass[preprint]{aastex}
\usepackage{underscore}
\usepackage{graphicx}
\begin{document}

\title{Cataclysmic Variables in the SUPERBLINK Proper Motion Survey\footnotemark[1]}
\footnotetext[1]{Based on Observations obtained at the MDM Observatory operated by Dartmouth College, Columbia University, The Ohio State University, and the University of Michigan}

\author{Julie N. Skinner\altaffilmark{2}, John R. Thorstensen\altaffilmark{2}}
\altaffiltext{2}{Dept. of Physics and Astronomy, 6127 Wilder Laboratory, Dartmouth College, Hanover, NH 03755-3528;\\
jns@dartmouth.edu}
\and
\author{S\'ebastien L\'epine\altaffilmark{3,4,5}}
\altaffiltext{3}{Department of Physics \& Astronomy, Georgia State University, 25 Park Place NE, Atlanta, GA, 30303}
\altaffiltext{4}{Department of Astrophysics, American Museum of Natural History, Central Park West at 79th Street, New York, NY 10024, USA}
\altaffiltext{5}{The Graduate Center, City University of New York, New York, NY}

\begin{abstract}
We have discovered a new high proper motion cataclysmic variable (CV) in the SUPERBLINK proper motion survey, which is sensitive to stars with proper motions greater than 40 mas yr$^{-1}$.  This CV was selected for follow-up observations as part of a larger search for CVs selected based on proper motions and their \textit{NUV-V} and \textit{V-K$_{s}$} colors.  We present spectroscopic observations from the 2.4m Hiltner Telescope at MDM Observatory.  The new CV's orbital period is near 96 minutes, its spectrum shows the double-peaked Balmer emission lines characteristic of quiescent dwarf novae, and its \textit{V} magnitude is near 18.2.  Additionally, we present a full list of known CVs in the SUPERBLINK catalog.
\end{abstract}

\section{Introduction}
Cataclysmic variable stars (CVs) are close, interacting binaries consisting of a white dwarf and a late-type dwarf star.  In these systems, the white dwarf accretes matter from the more extended secondary star through Roche-lobe overflow. \citet{Warner95} gives a comprehensive review of the field.  

Historically, there have been several avenues for the discovery of CVs.  The earliest CVs were discovered due to their brightening from an outburst (e.g. \citealt{Gaposchkin57}).  A few early novae were also discovered due to their emission lines \citep{Fleming12}.   Variability remains important to the discovery of new CVs.  The Catalina Real-Time Transient Survey (CRTS) repeatedly surveys 26,000 square degrees of the sky \citep{Drake09} and has now flagged over 750 objects as ``confirmed/likely" CVs discovered solely based on their variability.  However, discovery via variability preferentially selects systems that have an observed outburst.  \citet{Thor12} show that the CRTS also has a higher discovery rate of systems that have larger outburst amplitude.  Other optical variability surveys such as the Mobile Astronomical System of Telescope-Robots (MASTER; \citealt{Lipunov04}) and the Automated All-Sky Survey for SuperNovae (ASAS-SN; \citealt{Shappee13}) continue to find new CVs through variability.

X-ray emission from CVs is also an important discovery channel.  Specifically, the \textit{ROSAT} Bright Survey (RBS; \citealt{Schwope00}) contains bright, high Galactic latitude sources ($\vert$b$\vert$ $>$ 30$\degr$) culled as a flux-limited part of the \textit{ROSAT} All-Sky Survey \citep{Voges99}.  Including the CVs discovered from the deeper \textit{ROSAT} North Ecliptic Pole survey, this produced a flux-limited sample of 20 non-magnetic systems and 30 magnetic systems that were recently used to determine the space density and X-ray luminosity function of CVs \citep{Pretorius07b,Pretorius12,Pretorius13}.

Another successful avenue for the discovery of CVs relies on targeting emission-line objects or objects with characteristic colors within large optical surveys.  33 CVs were identified from the Palomar-Green Survey \citep{Ringwald93}, which covered 10700 deg$^{2}$ at high galactic latitude ($\vert$b$\vert$ $>$ 30$\degr$).  The PG survey identified objects based on their UV excess, selecting objects with \ub{} $<$ -0.46 with magnitudes of \textit{B} $>$ 16.16 \citep{Green86}.  The Hamburg Quasar Survey (HQS; \citet{Hagen95}) covered 13600 deg$^2$ with the 80cm Schmidt telescope at Calar Alto.  CVs were selected from their broad Balmer emission, their blue color, and variability \citep{Gansicke02}.  This resulted in the identification of 53 new CVs (see \citet{Aungwerojwit05} for an overview).  More recently, numerous CVs have been discovered in the Sloan Digital Sky Survey (SDSS; \citealt{Szkody03,Szkody04,Szkody05,Szkody06,Szkody07,Szkody09,Szkody11,Gansicke09}).  In particular, \citet{Gansicke09} found a number of intrinsically faint systems in SDSS, many of which have periods near the predicted period-minimum spike \citep{Kolb99}.  CVs in SDSS were identified through many methods \citep{Szkody02}.  Photometric selection criteria were developed, but identified far more detached white dwarf-M dwarf binaries (WD+dM).  Many CVs were detected because of their overlapping colors with some QSOs that were primary targets in SDSS.  Still others were detected in the spectroscopic survey as serendipitous targets.   While SDSS has provided the largest and deepest homogeneous sample of CVs to date, it still suffers from selection effects (e.g. \citet{Wils10, Thor12}).

All of these discovery methods have produced a large sample of CVs to study.  However, they all have selection biases and leave out some fraction of the population of CVs. \citet{Gansicke05} gives a more complete look at the surveys where CVs have been found and how selection biases have affected each sample.  Particularly, the faintest CVs with no known outbursts could easily go unnoticed.  Even with the careful work of \citet{Pretorius07b,Pretorius12,Pretorius13}, the space density of CVs remains uncertain.  Ideally, we would like to have a sample of CVs that would enable us to put an upper limit on their space density.  By using a survey of stars with large proper motions that is itself complete, we can begin to build a complete sample of the nearest CVs.

A few CVs have been discovered through their high proper motions (see Table \ref{hpmcvs}). GD552 was discovered in the Lowell Observatory Proper Motion Survey \citep{Giclas70} and later identified to be a CV \citep{Greenstein78} with an orbital period near 103 minutes \citep{Hessman90}.  More recently, \citet{UndaSanzana08} studied this object and suggested GD552 could have a brown dwarf companion.  They also provided evidence that this system is a `period bouncer' (post period minimum CV).  GP Com, also discovered in the Lowell Observatory Proper Motion Survey \citep{Giclas61}, is a double-degenerate helium CV (AM CVn type) with a 46.5 minute period \citep{Marsh99}.  V396 Hya was discovered as a high proper motion star in the Cal\'{a}n-ESO survey \citep{Ruiz01a} and is also an AM CVn type system \citep{Ruiz01bb}. With an orbital period near 65 minutes \citep{Thor08}, it is one of the longest period double degenerate helium CVs known.

This paper presents the first results from our systematic search for nearby CVs in the SUPERBLINK proper motion survey (see Section \ref{sec:Discovery}) .  We concentrate here on a single newly discovered CV, PM I03338+3320 and present a full list of the known CVs in the SUPERBLINK catalog.

\section{Discovery}
\label{sec:Discovery}
\begin{figure}
\figurenum{1}
\includegraphics[angle = 0,width={\columnwidth}]{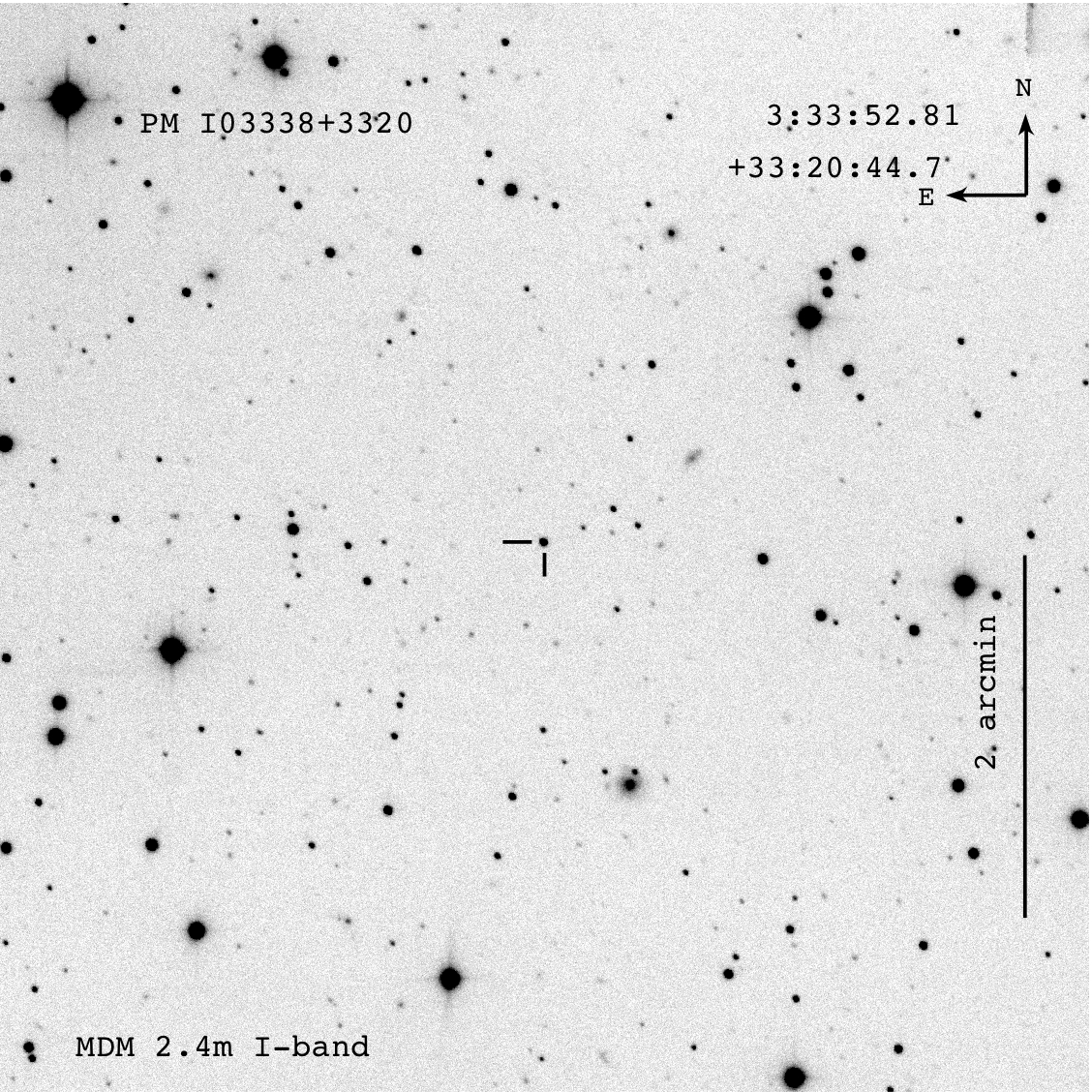}
\caption{A finding chart for PM I03338+3320 constructed from nine I-band images taken at the 2.4m Hiltner Telescope in mediocre seeing conditions.}
\label{chart}
\end{figure}

The identification of PM I03338+3320 as a CV was part of a larger project to search for nearby CVs in the SUPERBLINK proper motion survey. Astrometric and photometric properties are provided in Table \ref{03338props}.  A finding chart is also provided (Figure \ref{chart}).  SUPERBLINK is an all-sky survey with a proper motion limit of $\mu >$ 40 mas yr$^{-1}$.  It is based on an automated search of the Digitized Sky Surveys (DSS) using an image-differencing algorithm, followed by visual examination of candidate objects.  A description of the algorithm and quality control practices can be found in \citet{LSPM02} and \citet{LSPM05}.  New stars with proper motions larger than 450 mas yr$^{-1}$ can be found in \citet{LSPM02}, \citet{LSPM03}, and \citet{LSPM08}.  A full list of northern sky objects with proper motions larger than 150 mas yr$^{-1}$ can be found in \citet{LSPM05}, and an all-sky catalog of bright M dwarfs identified in the survey can be found in \citet{Lepine11}.  The current SUPERBLINK catalog is estimated to be $>$90 \% complete overall and extends to \textit{V} $\simeq$ 20.0.  It is more complete at higher galactic latitudes ($>$95\%) where crowding is less of a problem, although the low-latitude completeness is still estimated to be $>$80-85\%.  Analysis is complete for the northern sky, but is ongoing for parts of the southern sky.  SUPERBLINK had produced a catalog of 2,270,481 stars as of July 2011, which is the version we used for the current study.

\begin{figure}
\figurenum{2}
\centering
\includegraphics[angle= 0,width=0.8\textwidth]{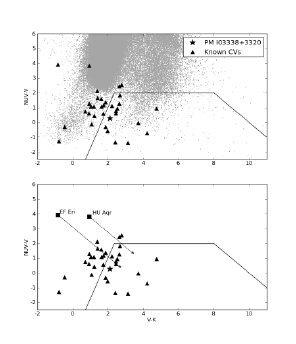}
\caption{\small{Top Panel: We show the position of PM I03338+3320 (solid star) on our color-color diagram along with the recovered known CVs (solid triangles) in SUPERBLINK.  For reference, the grey dots show all the objects in SUPERBLINK in this color-space.  The region shown within the solid lines illustrates our initial color selection used to isolate CVs and related systems.
Bottom Panel:  The two CVs marked as solid squares have \textit{V} magnitudes in SUPERBLINK consistent with being in an outburst or high state.  The endpoints of the arrows show their colors using an estimate of the low state \textit{V} magnitude from CRTS \citep{Drake09} for EF Eri and HU Aqr.}}
\label{colorplot}
\end{figure}

To isolate CVs in color-space, we used the \textit{V} magnitudes derived from the DSS plates as described in \citet{LSPM05}, near-\textit{UV} magnitudes from the \textit{Galaxy Evolution Explorer} (\textit{GALEX}; \citealt{Martin05}), and \textit{K$_{s}$} magnitudes from the Two Micron All Sky Survey (2MASS; \citealt{2mass}).  The top panel of Figure \ref{colorplot} shows PM I03338+3320 within the color cut, as well as the known CVs recovered in SUPERBLINK that have photometry available in all three bands (34 objects; see Table \ref{knowncvs}). For reference, other stars in SUPERBLINK are shown as grey dots.  Visual magnitudes in SUPERBLINK have been derived from the photographic plates of the Palomar Sky Survey (see \citet{LSPM05} for a complete explanation).  These have a typical uncertainty of 0.5 magnitudes.  Our initial color selection, shown as solid lines in both panels, was as follows. 
\begin{eqnarray}
\nonumber NUV-V < (2.75 * (V-K_{s})) - 4.5
\\\nonumber NUV-V \leq 2
\\\nonumber  NUV-V < 10 -  (V-K_{s})
\end{eqnarray}
For objects with no previous spectroscopic identification, we have obtained identifying spectra of 99\% of objects with declination $>$ -20$^{\circ}$ and 100\% of the northern sky objects with \textit{V} $<$ 17.0 within this color-selection (316 objects).

The bottom panel of Figure \ref{colorplot} shows the same region of color-space, but highlights two previously discovered CVs that exhibit colors inconsistent with what we would expect for CVs.  These objects have \textit{V} band photometry taken by the CRTS.  The arrows in the bottom panel show the revised \textit{NUV-V} and \textit{V-K$_{s}$} colors based on an estimate of the mean \textit{V} magnitude from either the Catalina Sky Survey or Siding Spring Survey light curves for these objects.  Including these two objects, this initial color-selection only includes 47\% of the known CVs in SUPERBLINK.  In the future, we plan to search the lower values of the \textit{V-K$_{s}$} color for undiscovered CVs.

\section{PM I03338+3320: A New High Proper Motion CV}
\subsection{Observations}
\label{sec:Observations}
We originally discovered PM I03338+3320 to be a CV in an exploratory spectroscopic run in September 2010 using the OSMOS spectrograph on the 2.4m Hiltner telescope at MDM observatory.   OSMOS \citep{Martini11}is a wide field imager and multi-object spectrograph.  We used OSMOS in long-slit mode with the volume-phase-holographic grism disperser, which gives a resolution of 2 \AA~.  The center slit yielded a wavelength coverage of 3100-5900 \AA, and the inner slit gives a wavelength coverage of 3960-6875 \AA.  

We obtained follow-up spectroscopy  in November 2010 with the same telescope using the modular spectrograph with $2048^{2}$ SITe CCD, which yielded a resolution of 3.5 \AA~and wavelength coverage from 4210-7520 \AA~ with vignetting toward the ends.  Exposures were typically 480 seconds in good seeing.  Wavelength calibrations were done using comparison lamp spectra taken during twilight, and shifts derived from the 5577 \AA~night sky line.  Reductions were done using standard IRAF routines except to extract one-dimensional spectra.  For this, we used a local implementation (developed by JT) of the optimal-extraction algorithm of \citet{Horne86}.  All radial velocity measurements were taken with this instrument.

\textit{BVRI} photometry was taken on January 24, 2011 with the 2.4m Hiltner telescope and a thick STIS 2048$^{2} $ CCD.  Landolt standards \citep{Landolt92} were used to calibrate.  This resulted in a \textit{V} magnitude of 18.2 and provided us with a check of the SUPERBLINK magnitude of \textit{V} = 18.03 $\pm$ 0.5.

\subsection{Discussion}
\label{sec:Discussion}

\begin{figure}
\figurenum{3}
\includegraphics[angle = -90,width={\columnwidth}]{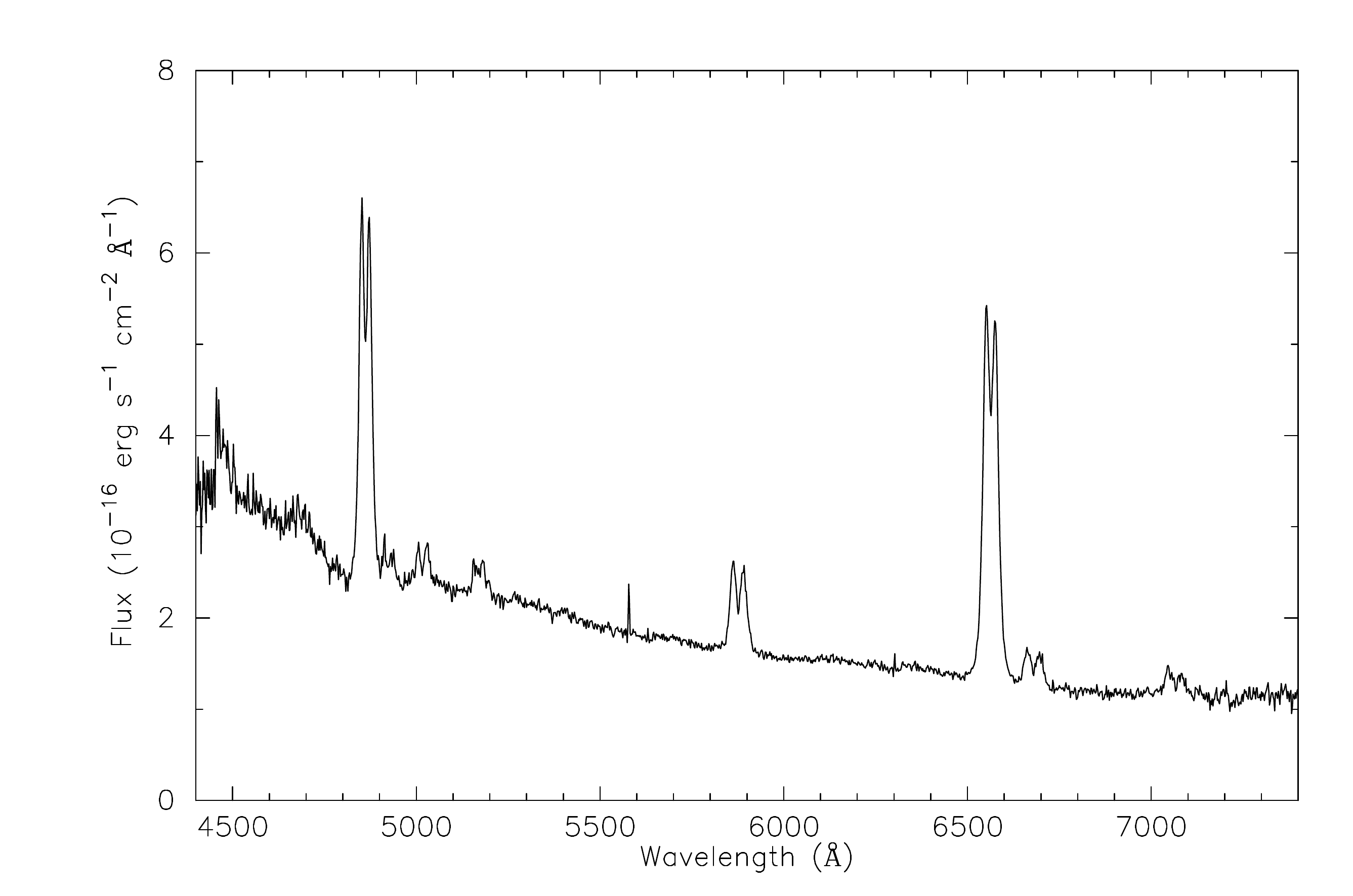}
\caption{The mean spectrum of PM I03338+3320, which shows double-peaked emission lines.  There is no visible contribution from a secondary.}
\label{avgspec}
\end{figure}

\begin{figure}
\figurenum{4}
\includegraphics[angle = 0,width={\columnwidth}]{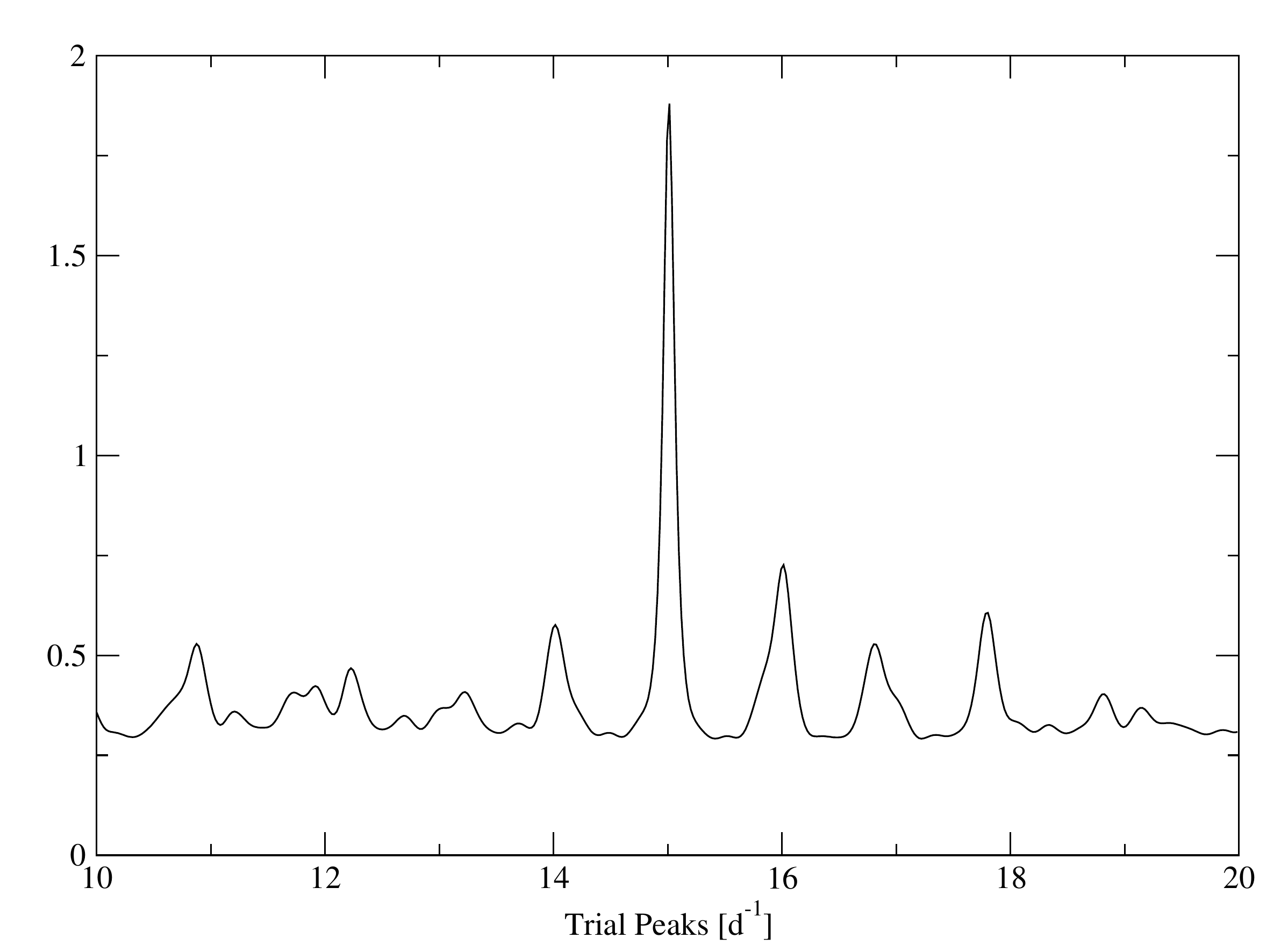}
\caption{The periodogram showing part of the range of periods tested.  The peak around 15 cycles per day stands out clearly as the best fit orbital period.}
\label{pgram}
\end{figure}

\begin{figure}
\figurenum{5}
\includegraphics[angle = -90,width={\columnwidth}]{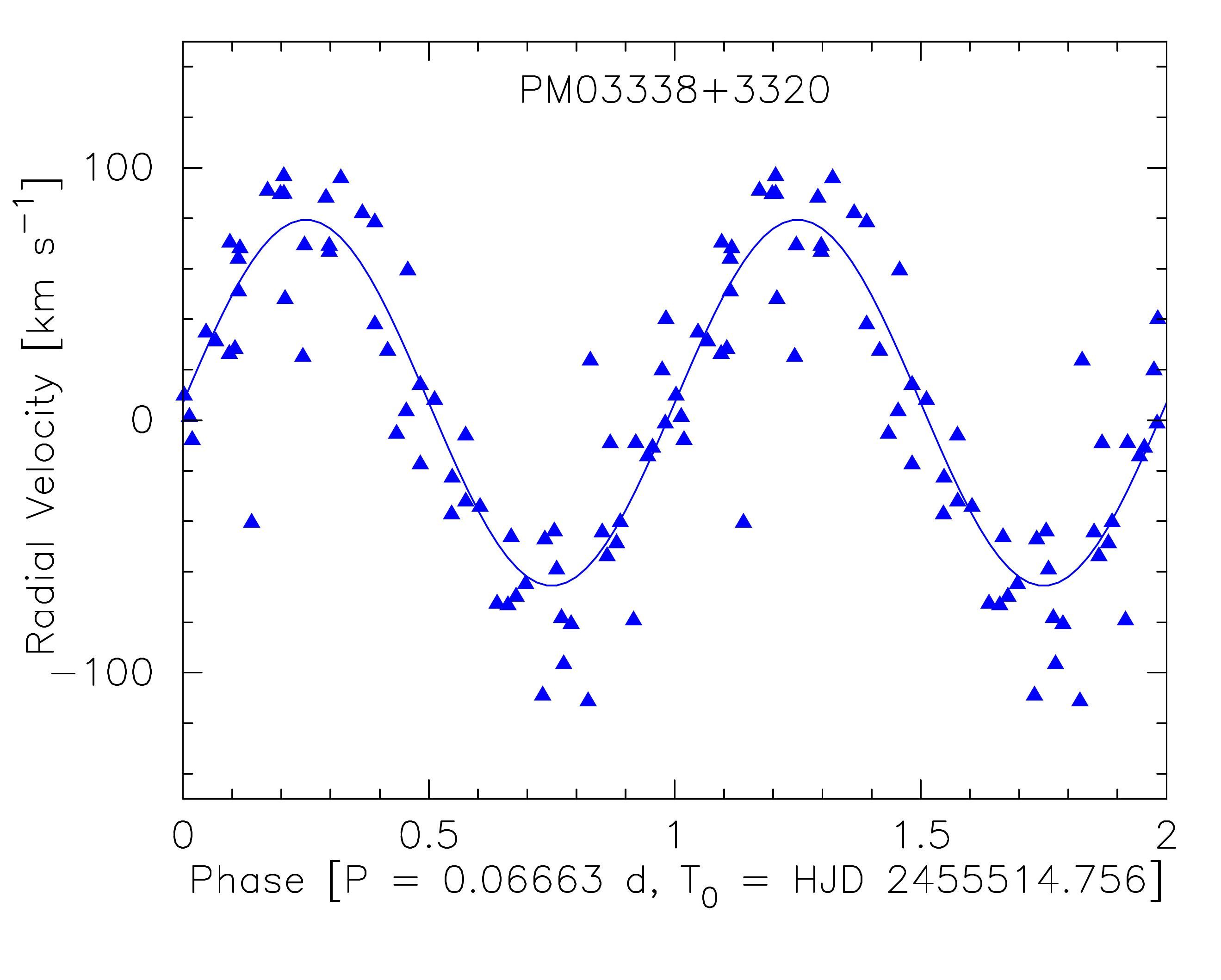}
\caption{Velocities from H$\alpha$ are shown by filled triangles along with the best-fit sinusoid as a solid line. The velocities are repeated for a second cycle for clarity.}
\label{rvcurve}
\end{figure}

The average spectrum shows double peaked emission lines throughout the orbit, but no visible contribution from a secondary star (Figure \ref{avgspec}).  Emission line velocities from H$\alpha$ were determined using a convolution method \citep{Schneider80,Shafter83} that uses an antisymmetric function with positive and negative Gaussians offset by an adjustable separation, $\alpha$.  This is then convolved with the observed line profile and the zero of the convolution is taken as the line center.  For PM I03338+3320, we adopted a value of $\alpha$ = 2284 km s$^{-1}$.  We used the ``residualgram'' method \citep{Thor96} to search for $P_{orb}$.  This search (Figure \ref{pgram}) yielded an unambiguous orbital period near 15 cycles per day or 96 minutes.  In Figure \ref{rvcurve}, we plot our best-fit sine curve along with the measured H$\alpha$ emission velocities.  The sine curve has the form $v(t) = \gamma + K \sin[2\pi (t - T_o)/P]$ and the fit parameters are given in Table \ref{03338sinefit}.

\begin{figure}
\figurenum{6}
\includegraphics[angle = 0,width={\columnwidth}]{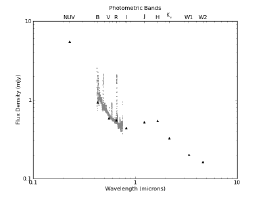}
\caption{The dereddened UV-optical-IR spectral energy distribution of PM I03338+3320 is shown.  The grey dots show the flux density from our mean spectrum.  The black triangles show the flux density in the corresponding photometric band shown on the top axis of the plot.}
\label{sed}
\end{figure}

PM I03338+3320 has no observed outbursts despite this region of the sky being monitored closely for transients.  In fact, there is a light curve for PMI03338+3320 in CRTS DR1 that shows variability around a mean magnitude of 17.5.  This means the CRTS survey should have detected any outburst from PM I03338+3320.  In addition, PM I03338+3320 has not been detected in x-ray surveys.  These things, along with its relatively faint \textit{V} magnitude, have contributed to it lurking undiscovered until now.  Figure \ref{sed} shows the UV-optical-IR spectral energy distribution for PM I03338+3320.  Extinction corrections were made using E(B-V) from \citet{Schlegel98} and the York Extinction Solver \citep{McCall04} with the galactic extinction law from \citet{Fitzpatrick99}.  Of interest is the large amount of flux in the H band, which may suggest a very late-type companion.  Near-IR spectroscopy of this object would be useful to determine the nature of the secondary star.

\section{Conclusion}
The identification of the high proper motion star PM I03338+3320 to be a CV is evidence that there is a portion of the nearby CV population yet to be discovered.  Our color-selection method proves to be promising with this discovery, but future work needs to be done probing bluer values of \textit{V-K$_{s}$} to develop a clear picture of the nearby population of CVs.  PM I03338+3320 is one of very few CVs to be discovered because of its high proper motion.  It has an orbital period of 0.06663 days with a spectrum that shows the double peaked spectral lines indicative of quiescent dwarf novae.  Its orbital period and lack of observed outburst makes it a likely SU UMa-type star. The collection of known CVs from SUPERBLINK presented here also gives us a large sample of systems with accurate proper motions to further help characterize the intrinsic population of CVs.

\section{Acknowledgements}
We gratefully acknowledge support from NSF grants AST 09-08419, AST-0708810, and AST-1008217.  This research was also made possible through the GALEX Guest Investigator program under NASA grant NNX09AF88G.  We thank the referee for helpful comments and suggestions.  We are grateful to the MDM staff who supported our observations.

\textit{GALEX} (\textit{Galaxy Evolution Explorer}) is a NASA Small Explorer, launched in 2003 April. We gratefully acknowledge NASA's support for construction, operation, and science analysis for the GALEX mission, developed in cooperation with the Centre National d'Etudes Spatiales of France and the Korean Ministry of Science and Technology.  

This publication makes use of data products from the Two Micron All Sky Survey, which is a joint project of the University of Massachusetts and the Infrared Processing and Analysis Center/California Institute of Technology, funded by the National Aeronautics and Space Administration and the National Science Foundation.

\bibliography{ms}

\begin{thebibliography}{}
\expandafter\ifx\csname natexlab\endcsname\relax\def\natexlab#1{#1}\fi

\bibitem[{Anderson {et~al.}(2005)Anderson, Haggard, Homer, Joshi, Margon,
  Silvestri, Szkody, Wolfe, Agol, Becker, Henden, Hall, Knapp, Richmond,
  Schneider, Stinson, Barentine, Brewington, Brinkmann, Harvanek, Kleinman,
  Krzesinski, Long, Neilsen, Nitta, \& Snedden}]{Anderson05}
Anderson, S.~F., Haggard, D., Homer, L., {et~al.} 2005, The Astronomical
  Journal, 130, 2230

\bibitem[{Appenzeller {et~al.}(1998)Appenzeller, Thiering, Zickgraf, Krautter,
  Voges, Chavarria, Kneer, Mujica, Pakull, Rosso, Ruzicka, Serrano, \&
  Ziegler}]{Appenzeller98}
Appenzeller, I., Thiering, I., Zickgraf, F.~J., {et~al.} 1998, The
  Astrophysical Journal Supplement Series, 117, 319

\bibitem[{Aungwerojwit {et~al.}(2005)Aungwerojwit, G{\"a}nsicke,
  Rodr{\'\i}guez-Gil, Hagen, Harlaftis, Papadimitriou, Lehto, Araujo-Betancor,
  Heber, Fried, Engels, \& Katajainen}]{Aungwerojwit05}
Aungwerojwit, A., G{\"a}nsicke, B.~T., Rodr{\'\i}guez-Gil, P., {et~al.} 2005,
  Astronomy and Astrophysics, 443, 995

\bibitem[{Baptista {et~al.}(1998)Baptista, Catalan, Horne, \&
  Zilli}]{Baptista98}
Baptista, R., Catalan, M.~S., Horne, K., \& Zilli, D. 1998, Monthly Notices of
  the Royal Astronomical Society, 300, 233

\bibitem[{Breedt {et~al.}(2012)Breedt, G{\"a}nsicke, Girven, Drake,
  Copperwheat, Parsons, \& Marsh}]{Breedt12}
Breedt, E., G{\"a}nsicke, B.~T., Girven, J., {et~al.} 2012, Monthly Notices of
  the Royal Astronomical Society, 423, 1437

\bibitem[{Cannizzo \& Mattei(1992)}]{Cannizzo92}
Cannizzo, J.~K., \& Mattei, J.~A. 1992, Astrophysical Journal, 401, 642

\bibitem[{Chen {et~al.}(2001)Chen, O'Donoghue, Stobie, Kilkenny, \&
  Warner}]{Chen01}
Chen, A., O'Donoghue, D., Stobie, R.~S., Kilkenny, D., \& Warner, B. 2001,
  Monthly Notices of the Royal Astronomical Society, 325, 89

\bibitem[{Cowley \& Crampton(1977)}]{Cowley77a}
Cowley, A.~P., \& Crampton, D. 1977, Astrophysical Journal, 212, L121

\bibitem[{Cowley {et~al.}(1977)Cowley, Crampton, \& Hesser}]{Cowley77b}
Cowley, A.~P., Crampton, D., \& Hesser, J.~E. 1977, Astrophysical Journal, 214,
  471

\bibitem[{Denisenko {et~al.}(2012)Denisenko, Podvorotny, Balanutsa, Shurpakov,
  Tiurina, Gorbovskoy, Lipunov, Kornilov, Belinski, Shatskiy, Chazov,
  Kuznetsov, Zimnukhov, Krushinsky, Zalozhnih, Popov, Bourdanov, Punanova,
  Ivanov, Yazev, Budnev, Konstantinov, Chuvalaev, Poleshchuk, Gress,
  Parkhomenko, Tlatov, Dormidontov, Senik, Yurkov, Sergienko, Varda, Sinyakov,
  Shumkov, Levato, Saffe, Mallamaci, Lopez, \& Podest}]{Denisenko12a}
Denisenko, D., Podvorotny, P., Balanutsa, P., {et~al.} 2012, The Astronomer's
  Telegram, 4441, 1

\bibitem[{Dillon {et~al.}(2008)Dillon, G{\"a}nsicke, Aungwerojwit,
  Rodr{\'\i}guez-Gil, Marsh, Barros, Szkody, Brady, Krajci, \&
  Oksanen}]{Dillon08}
Dillon, M., G{\"a}nsicke, B.~T., Aungwerojwit, A., {et~al.} 2008, Monthly
  Notices of the Royal Astronomical Society, 386, 1568

\bibitem[{Downes {et~al.}(2001)Downes, Webbink, Shara, Ritter, Kolb, \&
  Duerbeck}]{Downes01}
Downes, R.~A., Webbink, R.~F., Shara, M.~M., {et~al.} 2001, Publications of the
  Astronomical Society of the Pacific, 113, 764

\bibitem[{Drake {et~al.}(2009)Drake, Djorgovski, Mahabal, Beshore, Larson,
  Graham, Williams, Christensen, Catelan, Boattini, Gibbs, Hill, \&
  Kowalski}]{Drake09}
Drake, A.~J., Djorgovski, S.~G., Mahabal, A., {et~al.} 2009, The Astrophysical
  Journal, 696, 870

\bibitem[{Faulkner {et~al.}(1972)Faulkner, Flannery, \& Warner}]{Faulkner72}
Faulkner, J., Flannery, B.~P., \& Warner, B. 1972, Astrophysical Journal, 175,
  L79

\bibitem[{Fiedler {et~al.}(1997)Fiedler, Barwig, \& Mantel}]{Fiedler97}
Fiedler, H., Barwig, H., \& Mantel, K.~H. 1997, Astronomy and Astrophysics,
  327, 173

\bibitem[{Fitzpatrick(1999)}]{Fitzpatrick99}
Fitzpatrick, E.~L. 1999, Publications of the Astronomical Society of the
  Pacific, 111, 63

\bibitem[{Fleming \& Pickering(1912)}]{Fleming12}
Fleming, W. P.~S., \& Pickering, E.~C. 1912, Annals of the Astronomical
  Observatory of Harvard College ; v. 56, 56, 165

\bibitem[{G{\"a}nsicke(2005)}]{Gansicke05}
G{\"a}nsicke, B.~T. 2005, The Astrophysics of Cataclysmic Variables and Related
  Objects, 330, 3

\bibitem[{G{\"a}nsicke {et~al.}(2002)G{\"a}nsicke, Hagen, \&
  Engels}]{Gansicke02}
G{\"a}nsicke, B.~T., Hagen, H.-J., \& Engels, D. 2002, The Physics of
  Cataclysmic Variables and Related Objects, 261, 190

\bibitem[{G{\"a}nsicke {et~al.}(2009)G{\"a}nsicke, Dillon, Southworth,
  Thorstensen, Rodr{\'\i}guez-Gil, Aungwerojwit, Marsh, Szkody, Barros,
  Casares, de~Martino, Groot, Hakala, Kolb, Littlefair, Mart{\'\i}nez-Pais,
  Nelemans, \& Schreiber}]{Gansicke09}
G{\"a}nsicke, B.~T., Dillon, M., Southworth, J., {et~al.} 2009, Monthly Notices
  of the Royal Astronomical Society, 397, 2170

\bibitem[{Gaposchkin(1957)}]{Gaposchkin57}
Gaposchkin, C. H.~P. 1957, Amsterdam, North-Holland Pub. Co.; New York,
  Interscience Publishers, 1957.

\bibitem[{Garrison {et~al.}(1984)Garrison, Schild, Hiltner, \&
  Krzeminski}]{Garrison84}
Garrison, R.~F., Schild, R.~E., Hiltner, W.~A., \& Krzeminski, W. 1984,
  Astrophysical Journal, 276, L13

\bibitem[{Giclas {et~al.}(1961)Giclas, Burnham, \& Thomas}]{Giclas61}
Giclas, H.~L., Burnham, R., \& Thomas, N.~G. 1961, Bulletin / Lowell
  Observatory ; no. 112, 5, 61

\bibitem[{Giclas {et~al.}(1970)Giclas, Burnham, \& Thomas}]{Giclas70}
---. 1970, Bulletin / Lowell Observatory ; no. 153, 7, 183

\bibitem[{Green {et~al.}(1986)Green, Schmidt, \& Liebert}]{Green86}
Green, R.~F., Schmidt, M., \& Liebert, J. 1986, Astrophysical Journal
  Supplement Series (ISSN 0067-0049), 61, 305

\bibitem[{Greenstein \& Giclas(1978)}]{Greenstein78}
Greenstein, J.~L., \& Giclas, H. 1978, Astronomical Society of the Pacific, 90,
  460

\bibitem[{Griffiths {et~al.}(1979)Griffiths, Chaisson, Ward, Blades, Wilson, \&
  Johnston}]{Griffiths79}
Griffiths, R.~E., Chaisson, L., Ward, M.~J., {et~al.} 1979, Astrophysical
  Journal, 232, L27

\bibitem[{Hagen {et~al.}(1995)Hagen, Groote, Engels, \& Reimers}]{Hagen95}
Hagen, H.-J., Groote, D., Engels, D., \& Reimers, D. 1995, Astronomy and
  Astrophysics Supplement, 111, 195

\bibitem[{Harvey {et~al.}(1995)Harvey, Skillman, Patterson, \&
  Ringwald}]{Harvey95}
Harvey, D., Skillman, D.~R., Patterson, J., \& Ringwald, F.~A. 1995,
  Publications of the Astronomical Society of the Pacific, 107, 551

\bibitem[{Hessman {et~al.}(1984)Hessman, Robinson, Nather, \&
  Zhang}]{Hessman90}
Hessman, F.~V., Robinson, E.~L., Nather, R.~E., \& Zhang, E.-H. 1984,
  Astrophysical Journal, 286, 747

\bibitem[{Horne(1986)}]{Horne86}
Horne, K. 1986, Astronomical Society of the Pacific, 98, 609

\bibitem[{Howell \& Szkody(1988)}]{Howell88}
Howell, S., \& Szkody, P. 1988, Astronomical Society of the Pacific, 100, 224

\bibitem[{Joy(1954)}]{Joy54}
Joy, A.~H. 1954, Astrophysical Journal, 120, 377

\bibitem[{Kato {et~al.}(2009)Kato, Imada, Uemura, Nogami, Maehara, Ishioka,
  Baba, Matsumoto, Iwamatsu, Kubota, Sugiyasu, Soejima, Moritani, Ohshima,
  Ohashi, Tanaka, Sasada, Arai, Nakajima, Kiyota, Tanabe, Imamura, Kunitomi,
  Kunihiro, Taguchi, Koizumi, Yamada, Nishi, Kida, Tanaka, Ueoka, Yasui,
  Maruoka, Henden, Oksanen, Moilanen, Tikkanen, Aho, Monard, Itoh, Dubovsky,
  Kudzej, Dancikova, Vanmunster, Pietz, Bolt, Boyd, Nelson, Krajci, Cook,
  Torii, Starkey, Shears, Jensen, Masi, Hynek, {Nov{\'a}}, K, {Koci{\'a}}, N,
  {Kr{\'a}}, L, {Ku{\v c}{\'a}}, Kov{\'a}, Kolasa, {\v S}tastn{\'y}, Staels,
  Miller, Sano, de~Ponthi{\`e}re, Miyashita, Crawford, Brady, Santallo,
  Richards, Martin, Buczynski, Richmond, Kern, Davis, Crabtree, Beaulieu,
  Davis, Aggleton, Morelle, Pavlenko, Andreev, Baklanov, Koppelman, Billings,
  Urbancok, Ogmen, Heathcote, Gomez, Voloshina, Retter, Mularczyk, Zoczewski,
  Olech, Kedzierski, Pickard, Stockdale, Virtanen, Morikawa, Hambsch, Garradd,
  Gualdoni, Geary, Omodaka, Sakai, Michel, C{\'a}rdenas, Gazeas, Niarchos,
  Yushchenko, Mallia, Fiaschi, Good, Walker, James, Douzu, Julian, Butterworth,
  Shugarov, Volkov, Chochol, Katysheva, Rosenbush, Khramtsova, Kehusmaa,
  Reszelski, Bedient, Liller, Pojma{\'n}ski, Simonsen, Stubbings, Schmeer,
  Muyllaert, Kinnunen, Poyner, Ripero, \& Kriebel}]{Kato09}
Kato, T., Imada, A., Uemura, M., {et~al.} 2009, Publications of the
  Astronomical Society of Japan, 61, 395

\bibitem[{Kato {et~al.}(2013)Kato, Hambsch, Maehara, Masi, Nocentini, Dubovsky,
  Kudzej, Imamura, Ogi, Tanabe, Akazawa, Krajci, Miller, de~Miguel, Henden,
  Littlefield, Noguchi, Ishibashi, Ono, Kawabata, Kobayashi, Sakai, Nishino,
  Furukawa, Masumoto, Matsumoto, Ohshima, Nakata, Honda, Kinugasa, Hashimoto,
  Stein, Pickard, Kiyota, Pavlenko, Antonyuk, Baklanov, Antonyuk, Samsonov,
  Pit, Sosnovskij, Oksanen, Harlingten, Tyyska, Monard, Shugarov, Chochol,
  Kasai, Maeda, Hirosawa, Itoh, Sabo, Ulowetz, Morelle, Michel, Suarez, James,
  Dvorak, Voloshina, Richmond, Staels, Boyd, Andreev, Parakhin, Katysheva,
  Miyashita, Nakajima, Bolt, Padovan, Nelson, Starkey, Buczynski, Starr, Goff,
  Denisenko, Kochanek, Shappee, Stanek, Prieto, Itagaki, Kaneko, Stubbings,
  Muyllaert, Shears, Schmeer, Poyner, \& Marco}]{Kato13}
Kato, T., Hambsch, F.-J., Maehara, H., {et~al.} 2013, arXiv.org, 7069

\bibitem[{Kolb \& Baraffe(1999)}]{Kolb99}
Kolb, U., \& Baraffe, I. 1999, Monthly Notices, 309, 1034

\bibitem[{Kraft(1962)}]{Kraft62}
Kraft, R.~P. 1962, Astrophysical Journal, 135, 408

\bibitem[{Krzeminski(1965)}]{Krzeminski65}
Krzeminski, W. 1965, Astrophysical Journal, 142, 1051

\bibitem[{Krzeminski \& Serkowski(1977)}]{Krzeminski77}
Krzeminski, W., \& Serkowski, K. 1977, Astrophysical Journal, 216, L45

\bibitem[{Landolt(1992)}]{Landolt92}
Landolt, A.~U. 1992, Astronomical Journal (ISSN 0004-6256), 104, 340

\bibitem[{L{\'e}pine(2008)}]{LSPM08}
L{\'e}pine, S. 2008, The Astronomical Journal, 135, 2177

\bibitem[{L{\'e}pine \& Gaidos(2011)}]{Lepine11}
L{\'e}pine, S., \& Gaidos, E. 2011, The Astronomical Journal, 142, 138

\bibitem[{L{\'e}pine \& Shara(2005)}]{LSPM05}
L{\'e}pine, S., \& Shara, M.~M. 2005, The Astronomical Journal, 129, 1483

\bibitem[{L{\'e}pine {et~al.}(2002)L{\'e}pine, Shara, \& Rich}]{LSPM02}
L{\'e}pine, S., Shara, M.~M., \& Rich, R.~M. 2002, The Astronomical Journal,
  124, 1190

\bibitem[{L{\'e}pine {et~al.}(2003)L{\'e}pine, Shara, \& Rich}]{LSPM03}
---. 2003, The Astronomical Journal, 126, 921

\bibitem[{Lipunov {et~al.}(2004)Lipunov, Krylov, Kornilov, Borisov, Kuvshinov,
  Belinsky, Kuznetsov, Potanin, Antipov, Tyurina, Gorbovskoy, \&
  Chilingaryan}]{Lipunov04}
Lipunov, V.~M., Krylov, A.~V., Kornilov, V.~G., {et~al.} 2004, Astronomische
  Nachrichten, 325, 580

\bibitem[{Littlefair {et~al.}(2008)Littlefair, Dhillon, Marsh, G{\"a}nsicke,
  Southworth, Baraffe, Watson, \& Copperwheat}]{Littlefair08}
Littlefair, S.~P., Dhillon, V.~S., Marsh, T.~R., {et~al.} 2008, Monthly Notices
  of the Royal Astronomical Society, 388, 1582

\bibitem[{Marsh(1999)}]{Marsh99}
Marsh, T.~R. 1999, Monthly Notices of the Royal Astronomical Society, 304, 443

\bibitem[{Martin {et~al.}(2005)Martin, Fanson, Schiminovich, Morrissey,
  Friedman, Barlow, Conrow, Grange, Jelinsky, Milliard, Siegmund, Bianchi,
  Byun, Donas, Forster, Heckman, Lee, Madore, Malina, Neff, Rich, Small,
  Surber, Szalay, Welsh, \& Wyder}]{Martin05}
Martin, D.~C., Fanson, J., Schiminovich, D., {et~al.} 2005, The Astrophysical
  Journal, 619, L1

\bibitem[{Martini {et~al.}(2011)Martini, Stoll, Derwent, Zhelem, Atwood,
  Gonzalez, Mason, O'Brien, Pappalardo, Pogge, Ward, \& Wong}]{Martini11}
Martini, P., Stoll, R., Derwent, M.~A., {et~al.} 2011, Publications of the
  Astronomical Society of the Pacific, 123, 187

\bibitem[{McCall(2004)}]{McCall04}
McCall, M.~L. 2004, The Astronomical Journal, 128, 2144

\bibitem[{Mennickent {et~al.}(2001)Mennickent, Diaz, Skidmore, \&
  Sterken}]{Mennickent01}
Mennickent, R.~E., Diaz, M., Skidmore, W., \& Sterken, C. 2001, Astronomy and
  Astrophysics, 376, 448

\bibitem[{Nather {et~al.}(1981)Nather, Robinson, \& Stover}]{Nather81}
Nather, R.~E., Robinson, E.~L., \& Stover, R.~J. 1981, Astrophysical Journal,
  244, 269

\bibitem[{Nogami {et~al.}(1997)Nogami, Masuda, \& Kato}]{Nogami97}
Nogami, D., Masuda, S., \& Kato, T. 1997, Publications of the Astronomical
  Society of the Pacific, 109, 1114

\bibitem[{Page {et~al.}(2013)Page, Osborne, Beardmore, \& Schwarz}]{Page13}
Page, K.~L., Osborne, J.~P., Beardmore, A.~P., \& Schwarz, G.~J. 2013, The
  Astronomer's Telegram, 4920, 1

\bibitem[{Parisi {et~al.}(2014)Parisi, Masetti, Rojas, Jim{\'e}nez-Bail{\'o}n,
  Chavushyan, Palazzi, Bassani, Bazzano, Bird, Galaz, Minniti, Morelli, \&
  Ubertini}]{Parisi14}
Parisi, P., Masetti, N., Rojas, A.~F., {et~al.} 2014, Astronomy and
  Astrophysics, 561, A67

\bibitem[{Patterson(1979)}]{Patterson79}
Patterson, J. 1979, Astronomical Journal, 84, 804

\bibitem[{Patterson {et~al.}(1993)Patterson, Bond, Grauer, Shafter, \&
  Mattei}]{Patterson93}
Patterson, J., Bond, H.~E., Grauer, A.~D., Shafter, A.~W., \& Mattei, J.~A.
  1993, Astronomical Society of the Pacific, 105, 69

\bibitem[{Patterson {et~al.}(2005)Patterson, Thorstensen, Armstrong, Henden, \&
  Hynes}]{Patterson05}
Patterson, J., Thorstensen, J.~R., Armstrong, E., Henden, A.~A., \& Hynes,
  R.~I. 2005, Publications of the Astronomical Society of the Pacific, 117, 922

\bibitem[{Patterson {et~al.}(2003)Patterson, Thorstensen, Kemp, Skillman,
  Vanmunster, Harvey, Fried, Jensen, Cook, Rea, Monard, McCormick, Velthuis,
  Walker, Martin, Bolt, Pavlenko, O'Donoghue, Gunn, Nov{\'a}k, Masi, Garradd,
  Butterworth, Krajci, Foote, \& Beshore}]{Patterson03}
Patterson, J., Thorstensen, J.~R., Kemp, J., {et~al.} 2003, Publications of the
  Astronomical Society of the Pacific, 115, 1308

\bibitem[{Pretorius \& Knigge(2012)}]{Pretorius12}
Pretorius, M.~L., \& Knigge, C. 2012, Monthly Notices of the Royal Astronomical
  Society, 419, 1442

\bibitem[{Pretorius {et~al.}(2007)Pretorius, Knigge, O'Donoghue, Henry, Gioia,
  \& Mullis}]{Pretorius07b}
Pretorius, M.~L., Knigge, C., O'Donoghue, D., {et~al.} 2007, Monthly Notices of
  the Royal Astronomical Society, 382, 1279

\bibitem[{Pretorius {et~al.}(2013)Pretorius, Knigge, \& Schwope}]{Pretorius13}
Pretorius, M.~L., Knigge, C., \& Schwope, A.~D. 2013, Monthly Notices of the
  Royal Astronomical Society, 432, 570

\bibitem[{Pretorius {et~al.}(2004)Pretorius, Woudt, Warner, Bolt, Patterson, \&
  Armstrong}]{Pretorius04}
Pretorius, M.~L., Woudt, P.~A., Warner, B., {et~al.} 2004, Monthly Notices of
  the Royal Astronomical Society, 352, 1056

\bibitem[{Priedhorsky(1977)}]{Priedhorsky77}
Priedhorsky, W.~C. 1977, Astrophysical Journal, 212, L117

\bibitem[{Ramsay {et~al.}(2009)Ramsay, Rosen, Hakala, \& Barclay}]{Ramsay09}
Ramsay, G., Rosen, S., Hakala, P., \& Barclay, T. 2009, Monthly Notices of the
  Royal Astronomical Society, 395, 416

\bibitem[{Reinsch {et~al.}(1994)Reinsch, Burwitz, Beuermann, Schwope, \&
  Thomas}]{Reinsch94}
Reinsch, K., Burwitz, V., Beuermann, K., Schwope, A.~D., \& Thomas, H.~C. 1994,
  Astronomy and Astrophysics (ISSN 0004-6361), 291, L27

\bibitem[{Remillard {et~al.}(1994)Remillard, Schachter, Silber, \&
  Slane}]{Remillard94}
Remillard, R.~A., Schachter, J.~F., Silber, A.~D., \& Slane, P. 1994,
  Astrophysical Journal, 426, 288

\bibitem[{Ringwald(1993)}]{Ringwald93}
Ringwald, F.~A. 1993, PhD thesis, Dartmouth Coll., Hanover, NH.

\bibitem[{Ruiz {et~al.}(2001{\natexlab{a}})Ruiz, Rojo, Garay, \&
  Maza}]{Ruiz01bb}
Ruiz, M.~T., Rojo, P.~M., Garay, G., \& Maza, J. 2001{\natexlab{a}}, The
  Astrophysical Journal, 552, 679

\bibitem[{Ruiz {et~al.}(2001{\natexlab{b}})Ruiz, Wischnjewsky, Rojo, \&
  Gonzalez}]{Ruiz01a}
Ruiz, M.~T., Wischnjewsky, M., Rojo, P.~M., \& Gonzalez, L.~E.
  2001{\natexlab{b}}, The Astrophysical Journal Supplement Series, 133, 119

\bibitem[{Scaringi {et~al.}(2012)Scaringi, Groot, Verbeek, Greiss, Knigge, \&
  Kording}]{Scaringi13}
Scaringi, S., Groot, P.~J., Verbeek, K., {et~al.} 2012, Monthly Notices of the
  Royal Astronomical Society, 428, 2207

\bibitem[{Schachter {et~al.}(1996)Schachter, Remillard, Saar, Favata,
  Sciortino, \& Barbera}]{Schachter96}
Schachter, J.~F., Remillard, R., Saar, S.~H., {et~al.} 1996, Astrophysical
  Journal v.463, 463, 747

\bibitem[{Schlegel {et~al.}(1998)Schlegel, Finkbeiner, \& Davis}]{Schlegel98}
Schlegel, D.~J., Finkbeiner, D.~P., \& Davis, M. 1998, Astrophysical Journal
  v.500, 500, 525

\bibitem[{Schmidt {et~al.}(2005)Schmidt, Szkody, Homer, Smith, Chen, Henden,
  Solheim, Wolfe, \& Greimel}]{Schmidt05}
Schmidt, G.~D., Szkody, P., Homer, L., {et~al.} 2005, The Astrophysical
  Journal, 620, 422

\bibitem[{Schneider \& Young(1980)}]{Schneider80}
Schneider, D.~P., \& Young, P. 1980, Astrophysical Journal, 240, 871

\bibitem[{Schwope {et~al.}(2000)Schwope, Hasinger, Lehmann, Schwarz, Brunner,
  Neizvestny, Ugryumov, Balega, Tr{\"u}mper, \& Voges}]{Schwope00}
Schwope, A., Hasinger, G., Lehmann, I., {et~al.} 2000, Astronomische
  Nachrichten, 321, 1

\bibitem[{Schwope {et~al.}(2002)Schwope, Brunner, Buckley, Greiner, Heyden,
  Neizvestny, Potter, \& Schwarz}]{Schwope02}
Schwope, A.~D., Brunner, H., Buckley, D., {et~al.} 2002, Astronomy and
  Astrophysics, 396, 895

\bibitem[{Schwope {et~al.}(1999)Schwope, Schwarz, \& Greiner}]{Schwope99}
Schwope, A.~D., Schwarz, R., \& Greiner, J. 1999, Astronomy and Astrophysics,
  348, 861

\bibitem[{Schwope {et~al.}(1993)Schwope, Thomas, \& Beuermann}]{Schwope93}
Schwope, A.~D., Thomas, H.~C., \& Beuermann, K. 1993, Astronomy and
  Astrophysics, 271, L25

\bibitem[{Shafter(1983)}]{Shafter83}
Shafter, A.~W. 1983, Astrophysical Journal, 267, 222

\bibitem[{Shappee {et~al.}(2013)Shappee, Prieto, Grupe, Kochanek, Stanek,
  De~Rosa, Mathur, Zu, Peterson, Pogge, Komossa, Im, Jencson, Holoien, Basu,
  Beacom, Szczygiel, Brimacombe, Adams, Campillay, Choi, Contreras, Dietrich,
  Dubberley, Elphick, Foale, Giustini, Gonzalez, Hawkins, Howell, Hsiao, Koss,
  Leighly, Morrell, Mudd, Mullins, Nugent, Parrent, Phillips, Pojmanski,
  Rosing, Ross, Sand, Terndrup, Valenti, Walker, \& Yoon}]{Shappee13}
Shappee, B.~J., Prieto, J.~L., Grupe, D., {et~al.} 2013, arXiv.org, 1310.2241v2

\bibitem[{Sheets {et~al.}(2007)Sheets, Thorstensen, Peters, Kapusta, \&
  Taylor}]{Sheets07}
Sheets, H.~A., Thorstensen, J.~R., Peters, C.~J., Kapusta, A.~B., \& Taylor,
  C.~J. 2007, Publications of the Astronomical Society of the Pacific, 119, 494

\bibitem[{Skillman {et~al.}(2002)Skillman, Krajci, Beshore, Patterson, Kemp,
  Starkey, Oksanen, Vanmunster, Martin, \& Rea}]{Skillman02}
Skillman, D.~R., Krajci, T., Beshore, E., {et~al.} 2002, Publications of the
  Astronomical Society of the Pacific, 114, 630

\bibitem[{Skrutskie {et~al.}(2006)Skrutskie, Cutri, Stiening, Weinberg,
  Schneider, Carpenter, Beichman, Capps, Chester, Elias, Huchra, Liebert,
  Lonsdale, Monet, Price, Seitzer, Jarrett, Kirkpatrick, Gizis, Howard, Evans,
  Fowler, Fullmer, Hurt, Light, Kopan, Marsh, McCallon, Tam, Van~Dyk, \&
  Wheelock}]{2mass}
Skrutskie, M.~F., Cutri, R.~M., Stiening, R., {et~al.} 2006, The Astronomical
  Journal, 131, 1163

\bibitem[{Szkody \& Brownlee(1977)}]{Szkody77}
Szkody, P., \& Brownlee, D.~E. 1977, Astrophysical Journal, 212, L113

\bibitem[{Szkody \& Mattei(1984)}]{Szkody84}
Szkody, P., \& Mattei, J.~A. 1984, Astronomical Society of the Pacific, 96, 988

\bibitem[{Szkody {et~al.}(2002)Szkody, Anderson, Ag{\"u}eros, Covarrubias,
  Bentz, Hawley, Margon, Voges, Henden, Knapp, Vanden~Berk, Rest, Miknaitis,
  Magnier, Brinkmann, Csabai, Harvanek, Hindsley, Hennessy, Ivezic, Kleinman,
  Lamb, Long, Newman, Neilsen, Nichol, Nitta, Schneider, Snedden, \&
  York}]{Szkody02}
Szkody, P., Anderson, S.~F., Ag{\"u}eros, M., {et~al.} 2002, The Astronomical
  Journal, 123, 430

\bibitem[{Szkody {et~al.}(2003)Szkody, Fraser, Silvestri, Henden, Anderson,
  Frith, Lawton, Owens, Raymond, Schmidt, Wolfe, Bochanski, Covey, Harris,
  Hawley, Knapp, Margon, Voges, Walkowicz, Brinkmann, \& Lamb}]{Szkody03}
Szkody, P., Fraser, O., Silvestri, N., {et~al.} 2003, The Astronomical Journal,
  126, 1499

\bibitem[{Szkody {et~al.}(2004)Szkody, Henden, Fraser, Silvestri, Bochanski,
  Wolfe, Ag{\"u}eros, Warner, Woudt, Tramposch, Homer, Schmidt, Knapp,
  Anderson, Covey, Harris, Hawley, Schneider, Voges, \& Brinkmann}]{Szkody04}
Szkody, P., Henden, A., Fraser, O., {et~al.} 2004, The Astronomical Journal,
  128, 1882

\bibitem[{Szkody {et~al.}(2005)Szkody, Henden, Fraser, Silvestri, Schmidt,
  Bochanski, Wolfe, Ag{\"u}eros, Anderson, Mannikko, Downes, Schneider, \&
  Brinkmann}]{Szkody05}
Szkody, P., Henden, A., Fraser, O.~J., {et~al.} 2005, The Astronomical Journal,
  129, 2386

\bibitem[{Szkody {et~al.}(2006)Szkody, Henden, Ag{\"u}eros, Anderson,
  Bochanski, Knapp, Mannikko, Mukadam, Silvestri, Schmidt, Stephanik, Watson,
  West, Winget, Wolfe, Barentine, Brinkmann, Brewington, Downes, Harvanek,
  Kleinman, Krzesinski, Long, Neilsen, Nitta, Schneider, Snedden, \&
  Voges}]{Szkody06}
Szkody, P., Henden, A., Ag{\"u}eros, M., {et~al.} 2006, The Astronomical
  Journal, 131, 973

\bibitem[{Szkody {et~al.}(2007)Szkody, Henden, Mannikko, Mukadam, Schmidt,
  Bochanski, Ag{\"u}eros, Anderson, Silvestri, Dahab, Oguri, Schneider, Shin,
  Strauss, Knapp, \& West}]{Szkody07}
Szkody, P., Henden, A., Mannikko, L., {et~al.} 2007, The Astronomical Journal,
  134, 185

\bibitem[{Szkody {et~al.}(2009)Szkody, Anderson, Hayden, Kronberg, Mcgurk,
  Riecken, Schmidt, West, G{\"a}nsicke, Gomez-Moran, Schneider, Schreiber, \&
  Schwope}]{Szkody09}
Szkody, P., Anderson, S.~F., Hayden, M., {et~al.} 2009, The Astronomical
  Journal, 137, 4011

\bibitem[{Szkody {et~al.}(2011)Szkody, Anderson, Brooks, G{\"a}nsicke,
  Kronberg, Riecken, Ross, Schmidt, Schneider, Ag{\"u}eros, Gomez-Moran, Knapp,
  Schreiber, \& Schwope}]{Szkody11}
Szkody, P., Anderson, S.~F., Brooks, K., {et~al.} 2011, The Astronomical
  Journal, 142, 181

\bibitem[{Tapia(1977)}]{Tapia77}
Tapia, S. 1977, Astrophysical Journal, 212, L125

\bibitem[{Templeton {et~al.}(2006)Templeton, Leaman, Szkody, Henden, Cook,
  Starkey, Oksanen, Koppelman, Boyd, Nelson, Vanmunster, Pickard, Quinn,
  Huziak, Aho, James, Golovin, Pavlenko, Durkee, Crawford, Walker, \&
  P{\"a}{\"a}kk{\"o}nen}]{Templeton06}
Templeton, M.~R., Leaman, R., Szkody, P., {et~al.} 2006, Publications of the
  Astronomical Society of the Pacific, 118, 236

\bibitem[{Thorstensen(2003)}]{Thor03}
Thorstensen, J.~R. 2003, The Astronomical Journal, 126, 3017

\bibitem[{Thorstensen {et~al.}(2002)Thorstensen, Fenton, Patterson, Kemp,
  Krajci, \& Baraffe}]{Thor02}
Thorstensen, J.~R., Fenton, W.~H., Patterson, J.~O., {et~al.} 2002, The
  Astrophysical Journal, 567, L49

\bibitem[{Thorstensen {et~al.}(2006)Thorstensen, L{\'e}pine, \& Shara}]{Thor06}
Thorstensen, J.~R., L{\'e}pine, S., \& Shara, M. 2006, The Publications of the
  Astronomical Society of the Pacific, 118, 1238

\bibitem[{Thorstensen {et~al.}(2008)Thorstensen, L{\'e}pine, \& Shara}]{Thor08}
---. 2008, The Astronomical Journal, 136, 2107

\bibitem[{Thorstensen {et~al.}(1996)Thorstensen, Patterson, Shambrook, \&
  Thomas}]{Thor96}
Thorstensen, J.~R., Patterson, J.~O., Shambrook, A., \& Thomas, G. 1996,
  Publications of the Astronomical Society of the Pacific, 108, 73

\bibitem[{Thorstensen \& Skinner(2012)}]{Thor12}
Thorstensen, J.~R., \& Skinner, J.~N. 2012, The Astronomical Journal, 144, 81

\bibitem[{Uemura {et~al.}(2000)Uemura, Kato, Matsumoto, Takamizawa, Schmeer,
  Jensen, Vanmunster, Nov{\'a}k, Martin, Pietz, Buczynski, Kinnunen, Moilanen,
  Oksanen, Cook, Watanabe, Maehara, \& Itoh}]{Uemura00}
Uemura, M., Kato, T., Matsumoto, K., {et~al.} 2000, Publications of the
  Astronomical Society of Japan, 52, L9

\bibitem[{Unda-Sanzana {et~al.}(2008)Unda-Sanzana, Marsh, G{\"a}nsicke, Maxted,
  Morales-Rueda, Dhillon, Thoroughgood, Tremou, Watson, \&
  Hinojosa-Go{\~n}i}]{UndaSanzana08}
Unda-Sanzana, E., Marsh, T.~R., G{\"a}nsicke, B.~T., {et~al.} 2008, Monthly
  Notices of the Royal Astronomical Society, 388, 889

\bibitem[{Uthas {et~al.}(2011)Uthas, Knigge, Long, Patterson, \&
  Thorstensen}]{Uthas11}
Uthas, H., Knigge, C., Long, K.~S., Patterson, J., \& Thorstensen, J. 2011,
  Monthly Notices of the Royal Astronomical Society: Letters, 414, L85

\bibitem[{Voges {et~al.}(1999)Voges, Aschenbach, Boller, Br{\"a}uninger, Briel,
  Burkert, Dennerl, Englhauser, Gruber, Haberl, Hartner, Hasinger, K{\"u}rster,
  Pfeffermann, Pietsch, Predehl, Rosso, Schmitt, Tr{\"u}mper, \&
  Zimmermann}]{Voges99}
Voges, W., Aschenbach, B., Boller, T., {et~al.} 1999, Astronomy and
  Astrophysics, 349, 389

\bibitem[{Wagner {et~al.}(1988)Wagner, Sion, Liebert, \& Starrfield}]{Wagner88}
Wagner, R.~M., Sion, E.~M., Liebert, J., \& Starrfield, S.~G. 1988,
  Astrophysical Journal, 328, 213

\bibitem[{Warner(1995)}]{Warner95}
Warner, B. 1995, Camb. Astrophys. Ser., Vol. 28,

\bibitem[{Warner \& Nather(1972)}]{Warner72}
Warner, B., \& Nather, R.~E. 1972, Monthly Notices of the Royal Astronomical
  Society, 159, 429

\bibitem[{Wils {et~al.}(2010)Wils, G{\"a}nsicke, Drake, \& Southworth}]{Wils10}
Wils, P., G{\"a}nsicke, B.~T., Drake, A.~J., \& Southworth, J. 2010, Monthly
  Notices of the Royal Astronomical Society, 402, 436

\bibitem[{Witham {et~al.}(2007)Witham, Knigge, Aungwerojwit, Drew,
  G{\"a}nsicke, Greimel, Groot, Roelofs, Steeghs, \& Woudt}]{Witham07}
Witham, A.~R., Knigge, C., Aungwerojwit, A., {et~al.} 2007, Monthly Notices of
  the Royal Astronomical Society, 382, 1158

\bibitem[{Woudt {et~al.}(2005)Woudt, Warner, \& Spark}]{Woudt05}
Woudt, P.~A., Warner, B., \& Spark, M. 2005, Monthly Notices of the Royal
  Astronomical Society, 364, 107

\bibitem[{Zharikov {et~al.}(2006)Zharikov, Tovmassian, Napiwotzki, Michel, \&
  Neustroev}]{Zharikov06}
Zharikov, S.~V., Tovmassian, G.~H., Napiwotzki, R., Michel, R., \& Neustroev,
  V. 2006, Astronomy and Astrophysics, 449, 645

\end{thebibliography}

\begin{deluxetable}{cccc}
\tablenum{1}
\tabletypesize{\small}
\setlength{\tabcolsep}{0.07in}
\tablecolumns{4}
\tablewidth{0in}
\tablecaption{CVs Discovered due to their Proper Motion\label{hpmcvs}}
\tablehead{
\colhead{Star} &
\colhead{[$\mu_{\alpha}, \mu_{\delta}$]} &
\colhead{Distance} &
\colhead{Discovery Reference} \\
\colhead{} &
\colhead{(mas yr$^{-1}$)} &
\colhead{(pc)} &
\colhead{}
}
\startdata
GD 552 & 125,-40\tablenotemark{a}  & 74(3)\tablenotemark{a} & \citet{Greenstein78}\\
GP Com & -343, +32 & 68(+7,-6)\tablenotemark{b} & \citet{Marsh99}\\
V396 Hya & -277, -52 & 90(+12,-10)\tablenotemark{c} & \citet{Ruiz01a}\\
\enddata
\tablenotetext{a}{J. Thorstensen, private communication}
\tablenotetext{b}{\citet{Thor03}}
\tablenotetext{c}{\citet{Thor08}}
\end{deluxetable}

\begin{deluxetable}{ccccccccccc}
\rotate
\tablenum{2}
\tabletypesize{\small}
\setlength{\tabcolsep}{0.07in}
\tablecolumns{11}
\tablewidth{0in}
\tablecaption{Astrometric and Photometric Properties of PM I03338+3320\label{03338props}}
\tablehead{
\colhead{$\alpha$} &
\colhead{$\delta$} &
\colhead{$\mu$} &
\colhead{[$\mu_{\alpha}, \mu_{\delta}$]} &
\colhead{X-Ray} &
\colhead{FUV} &
\colhead{NUV} &
\colhead{J} &
\colhead{H} &
\colhead{K$_{s}$} &
\colhead{V} \\
\colhead{(J2000.0)} &
\colhead{(J2000.0)} &
\colhead{(mas yr$^{-1}$)} &
\colhead{(mas yr$^{-1}$)} &
\colhead{(counts s$^{-1}$)} &
\colhead{(mag)} &
\colhead{(mag)} &
\colhead{(mag)} &
\colhead{(mag)} &
\colhead{(mag)} &
\colhead{(mag)}
}
\startdata
53.469872 & 33.345783 & 52 & 52,-8 & $\cdots$ & $\cdots$ & 18.31 & 16.55 & 15.90 & 15.92 & 18.03 \\
\enddata
\tablecomments{For more detailed information on the quantities listed in this table, see \citet{Lepine11} and references therein.  Proper motion measurements from SUPERBLINK have a typical uncertainty of $\pm$ 8 mas yr$^{-1}$.  X-ray flux from the ROSAT all-sky point source catalog \citep{Voges99}.  Far- and Near-UV photometry from the \textit{GALEX} 5th data release \citep{Martin05}.  J, H, and K$_{s}$ photometry from the 2MASS catalog \citep{2mass}.  V magnitude determined as outlined in \citet{LSPM05}.}
\end{deluxetable}

\begin{deluxetable}{lcccccccc}
\tablenum{3}
\tabletypesize{\scriptsize}
\setlength{\tabcolsep}{0.07in}
\tablecolumns{9}
\tablewidth{0pt}
\tablecaption{Known CVs Recovered in SUPERBLINK\label{knowncvs}}
\tablehead{
\colhead{Name} &
\colhead{[$\mu_{\alpha}, \mu_{\delta}$]} &
\colhead{\textit{V}} &
\colhead{\textit{NUV-V}} &
\colhead{\textit{V-K$_{s}$}} &
\colhead{In} &
\colhead{Type} & 
\colhead{SUPERBLINK} &
\colhead{Reference} \\
\colhead{} &
\colhead{(mas yr$^{-1}$)} &
\colhead{} &
\colhead{} &
\colhead{} &
\colhead{Color-Selection?} &
\colhead{} &
\colhead{ID} &
\colhead{}
}
\startdata
  FL Psc & [-69,-18] & 16.78 & 1.09 & 1.2 & n & DN & PM I00251+1217 & 61\\ 
  J0032-7420 & [140,52] & 17.84 & -- & 5.4 & -- & N-EG & PM I00329-7420 & 32\\  
  EQ Cet & [29,-34] & 17.97 & 1.13 & 2.23 & y & P & PM I01288-2339 & 49\\  
  GZ Cet & [-38,-53] & 18.52 & -- & 3.22 & -- & DN & PM I01370-0912 &  38, 56\\  
  KT Per & [59,-9] & 15.22 & -- & 3.49 & -- & DN & PM I01371+5057 & 54\\  
  FL Cet & [49,18] & 18.62 & 0.61 & 2.45 & y & P & PM I01557+0028 &  46, 55\\  
  PQ And & [38,30] & 17.76 & -- & -- & -- & DN & PM I02294+4002 & 37\\
  EF Eri & [123,-47] & 14.53 & 3.94 & -0.84 & y* & P & PM I03142-2235 & 18\\  
  RBS 490 & [1,-105] & 17.47 & 0.58 & 1.73 & n & DN & PM I03541-1652 & 63\\
  IM Eri & [52,-29] & 11.92 & 0.74 & 0.71 & n & NL & PM I04246-2007 & 7\\  
  1RXS J042608.9+354151 & [17,-48] & 15.86 & 1.66 & 1.41 & n & DN & PM I04261+3541W &  10, 23\\
  BF Eri & [28,-109] & 15.23 & -- & 2.68 & -- & DN & PM I04394-0435 &  45, 51\\
  IPHAS J0528 & [34,-36] & 16.46 & -- & 1.1 & -- & P & PM I05285+2838 &  3, 69\\
  UW Pic & [28,69] & 16.51 & 1.59 & 1.62 & n & P & PM I05315-4624 & 41\\
  CSS100114:055843+000626 & [14,-41] & 19.06 & -- & 3.58 & -- & DN & PM I05587+0006 &  13, 64\\
  IR Gem & [52,-21] & 15.74 & 0.44 & 1.21 & n & DN & PM I06475+2806 & 54\\
  PBC J0706.7+0327 & [41,-30] & 16.34 & -- & 1.8 & -- & IP & PM I07068+0324 & 33\\
  U Gem & [-26,-32] & 14.54 & -0.04 & 3.71 & y & DN & PM I07550+2200 &  23, 25\\
  YZ Cnc & [38,-58] & 15.24 & -1.35 & 2.41 & y & DN & PM I08109+2808E & 34\\
  VV Pup & [19,-69] & 15.98 & -- & 1.44 & -- & P & PM I08151-1903 &  47, 60\\
  IX Vel & [-27,-96] & 9.44 & -- & 0.61 & -- & NL & PM I08153-4913 & 16\\
  LV Cnc & [-53,-4] & 17.7 & 1.01 & -- & -- & CV & PM I09197+0857 & 57\\
  IY UMa & [-42,-8] & 16.61 & 1.18 & 1.75 & n & DN & PM I10439+5807 & 65\\
  1RXS J105010.3-140431 & [-181,-4] & 16.76 & 1.09 & 1.02 & n & DN & PM I10501-1404 & 29\\
  SX LMi & [22,-34] & 16.3 & 0.62 & 0.91 & n & DN & PM I10545+3006W &  31, 67\\
  AN UMa & [-41,-20] & 16.94 & 1.37 & 1.87 & n & P & PM I11044+4503 & 26\\
  AR UMa & [-77,2] & 16.4 & -1.39 & 3.13 & y & P & PM I11157+4258 & 42\\
  RZ Leo & [-42,35] & 19.6 & -0.69 & 4.21 & y & DN & PM I11373+0148 & 20\\
  SDSS J113826.73+061919.5 & [-50,11] & 18.17 & 1.03 & -- & -- & CV & PM I11384+0619 & 58\\
  QZ Vir & [-99,-44] & 15.81 & -0.57 & 1.98 & y & DN & PM I11384+0322 & 24\\
  V1040 Cen & [-126,82] & 14.13 & -- & 1.19 & -- & DN & PM I11554-5641 & 36\\
  SDSS J121913.04+204938.3 & [21,-73] & 19.38 & 0.45 & -- & -- & CV & PM I12192+2049 & 58\\
  AM CVn & [34,24] & 14.19 & -0.29 & -0.46 & n & AC & PM I12349+3737 & 14\\
  V406 Vir & [-135,-36] & 18.29 & -0.31 & 1.87 & y & DN & PM I12382-0339 &  56, 71\\
  SDSS J125044.42+154957.3 & [-72,-59] & 18.28 & 0.94 & 2.52 & y & P & PM I12507+1549 & 5\\
  GP Com & [-333,47] & 16.22 & -0.12 & 1.07 & n & AC & PM I13057+1801 &  28, 30\\
  V396 Hya & [-257,-22] & 18.26 & 0.49 & -- & -- & AC & PM I13127-2321N & 43\\
  UX UMa & [-41,16] & 13.19 & 1.28 & 0.93 & n & NL & PM I13366+5154 & 68\\
  SDSS J141118.31+481257.6 & [-31,35] & 19.37 & 0.11 & -- & -- & AC & PM I14113+4812 & 2\\
  SDSS J143317.78+101123.3 & [-2,-55] & 18.86 & 0.47 & -- & -- & DN & PM I14332+1011 & 27\\
  NZ Boo & [-78,-31] & 17.59 & 0.79 & 2.44 & y & DN & PM I15026+3334 &  58, 27\\
  OV Boo & [-146,54] & 18.61 & -0.12 & -- & -- & CV & PM I15073+5230 & 66\\
  PP Boo & [-59,-23] & 19.45 & 1.04 & -- & -- & CV & PM I15142+4549 & 11\\
  SDSS J152212+0803 & [-58,18] & 15.89 & -- & 2.04 & -- & CV & PM I15222+0803W & 59\\
  SDSS J152419.33+220920.0 & [-44,23] & 19.32 & 0.86 & -- & -- & DN & PM I15243+2209 &  22, 59\\
  MLS120517:152507-032655 & [15,-55] & 18.22 & 1.83 & 2.67 & y & CV & PM I15251-0326 & 13\\
  V2051 Oph & [-18,-58] & 15.07 & -- & -- & -- & DN & PM I17083-2548S & 4\\
   EX Dra & [41,8] & 14.7 & 2.45 & 2.64 & n & DN & PM I18042+6754 & 15\\
  AM Her & [-43,23] & 13.63 & 1.27 & 2.63 & y & P & PM I18162+4952 &  8, 39, 53, 60\\
  KIS J192703.08+421131.7 & [49,80] & 19.05 & -- & -- & -- & CV & PM I19270+4211 & 44\\
  V3885 Sgr & [30,-40] & 10.33 & -- & 0.71 & -- & NL & PM I19476-4200 & 9\\
   V503 Cyg & [-18,-48] & 15.83 & -- & 0.63 & -- & DN & PM I20272+4341 & 19\\
  AE Aqr & [71,15] & 11.54 & 2.56 & 2.77 & n & IP & PM I20401-0052 & 21\\
  HU Aqr & [-72,-42] & 14.56 & 3.83 & 0.93 & y* & P & PM I21079-0517 & 48\\
  VY Aqr & [34,-32] & 16.23 & 1.08 & 1.64 & n & DN & PM I21121-0849 & 35\\
  SS Cyg & [107,30] & 11.69 & -- & 3.39 & -- & DN & PM I21427+4335 & 6\\
  RX J2237.5+0827 & [-92,-44] & 15.55 & -1.27 & -0.77 & n & CV: & PM I22375+0828 &  1, 12\\
  V367 Peg & [39,-20] & 18.38 & -- & 2.93 & -- & DN & PM I22450+1655 & 70\\
  2XMMiJ225036.9+573154 & [53,30] & 19.46 & -- & -- & -- & P & PM I22506+5731 & 40\\
  GD 552 & [114,-39] & 16.47 & -- & 2.01 & -- & DN & PM I22506+6328 & 17\\
  V405 Peg & [-40,-53] & 16.56 & 0.95 & 4.75 & y & DN & PM I23098+2135 & 50\\
  EI Psc & [-40,39] & 15.48 & 2.12 & 1.38 & n & DN & PM I23299+0628 &  52, 62\\
\enddata

\tablecomments{V magnitudes from SUPERBLINK catalog and have a typical uncertainty of 0.5 mag (see text).  EF Eri and HU Aqr have been included in the color selection based on their photometry from CRTS.  This is indicated by an asterisk next to their color selection status in column 6.  To the best of our ability, discovery and type references have been quoted for each object.  In the case of systems discovered in the past $\sim$25 years, this is straightforward.  For long-studied, well-known systems, these references become harder to pinpoint.  In some cases, we have referenced a more modern paper that gives the history of the study of the object.  Type Abbreviations: AC - AM CVn, CV - unclassified, DN - Dwarf Nova, EG - extragalactic, IP - Intermediate Polar, N - Nova, NL - Nova-like variable, P - Polar; A colon indicates uncertainty in classification.}

\tablerefs{(1) \citet{Appenzeller98}, (2) \citet{Anderson05}, (3) B. G\"{a}nsicke,  private communication, (4) \citet{Baptista98}, (5) \citet{Breedt12}, (6) \citet{Cannizzo92}, (7) \citet{Chen01}, (8) \citet{Cowley77a}, (9) \citet{Cowley77b}, (10) \citet{Denisenko12a}, (11) \citet{Dillon08}, (12) \citet{Downes01}, (13) \citet{Drake09}, (14) \citet{Faulkner72}, (15) \citet{Fiedler97}, (16) \citet{Garrison84}, (17) \citet{Greenstein78}, (18) \citet{Griffiths79}, (19) \citet{Harvey95}, (20) \citet{Howell88}, (21) \citet{Joy54}, (22) \citet{Kato09}, (23) \citet{Kato13}, (24) \citet{Kraft62}, (25) \citet{Krzeminski65}, (26) \citet{Krzeminski77}, (27) \citet{Littlefair08}, (28) \citet{Marsh99}, (29) \citet{Mennickent01}, (30) \citet{Nather81}, (31) \citet{Nogami97}, (32) \citet{Page13}, (33) \citet{Parisi14}, (34) \citet{Patterson79}, (35) \citet{Patterson93}, (36) \citet{Patterson03}, (37) \citet{Patterson05}, (38) \citet{Pretorius04}, (39) \citet{Priedhorsky77}, (40) \citet{Ramsay09}, (41) \citet{Reinsch94}, (42) \citet{Remillard94}, (43) \citet{Ruiz01bb}, (44) \citet{Scaringi13}, (45) \citet{Schachter96}, (46) \citet{Schmidt05}, (47) \citet{Schneider80}, (48) \citet{Schwope93}, (49) \citet{Schwope99}, (50) \citet{Schwope02}, (51) \citet{Sheets07}, (52) \citet{Skillman02}, (53) \citet{Szkody77}, (54) \citet{Szkody84}, (55) \citet{Szkody02}, (56) \citet{Szkody03}, (57) \citet{Szkody05}, (58) \citet{Szkody06}, (59) \citet{Szkody09}, (60) \citet{Tapia77}, (61) \citet{Templeton06}, (62) \citet{Thor02}, (63) \citet{Thor06}, (64) \citet{Thor12}, (65) \citet{Uemura00}, (66) \citet{Uthas11}, (67) \citet{Wagner88}, (68) \citet{Warner72}, (69) \citet{Witham07}, (70) \citet{Woudt05}, (71) \citet{Zharikov06}}
\end{deluxetable}

\begin{deluxetable}{ccccccc}
\tablenum{4}
\tabletypesize{\small}
\setlength{\tabcolsep}{0.07in}
\tablecolumns{7}
\tablewidth{0in}
\tablecaption{Sinusoidal Fit\label{03338sinefit}}
\tablehead{
\colhead{Data Set} &
\colhead{Epoch (days)} &
\colhead{Period (days)} &
\colhead{K (km s$^{-1}$)} &
\colhead{$\gamma$ (km s$^{-1}$)} &
\colhead{$\sigma$ (km s$^{-1}$)} &
\colhead{N}
}
\startdata
H$\alpha$ emission & 55514.7558(10) & 0.06663(7) &  73(7) & $ 7(5)$ & 64 &  22 \\
\enddata
\tablecomments{K is the amplitude of the sinusoidal fit, $\gamma$ is the mean velocity, and $\sigma$ is the error in a single measure based on the scatter around the best fit.}
\end{deluxetable}

\end{document}